\documentclass[11pt]{article}
\pdfoutput=1
\usepackage{graphicx}
\usepackage{hyperref}
\usepackage{epsfig}
\usepackage{amsmath}
\usepackage{amssymb}
\usepackage{multirow}
\usepackage{color}
\usepackage[nosort]{cite}
\usepackage{hyperref}
\usepackage[fontsize=11.4pt]{scrextend}
\usepackage{braket,mathrsfs,slashed}
\usepackage{setspace}
\usepackage[caption=false]{subfig}
\newlength{\abstractwidth}
\newcommand{\be}{\begin{equation}}
\newcommand{\ee}{\end{equation}}

\renewcommand{\title}[1]{\vbox{\center\bf{\Large{#1}}}\vspace{5mm}}
\renewcommand{\author}[1]{\vbox{\center#1}\vspace{5mm}}
\newcommand{\address}[1]{\vbox{\center\em#1}}
\newcommand{\email}[1]{\vbox{\center\tt#1}\vspace{5mm}}

\usepackage{mathrsfs} 

\usepackage{etoolbox} 
\usepackage{multirow}
\renewcommand\[{\begin{equation}}
\renewcommand\]{\end{equation}}

\newcommand{\ba}{\begin{eqnarray}}
\newcommand{\ea}{\end{eqnarray}}

\hypersetup{pdfstartview=FitV,colorlinks=true,linkcolor=midblue,citecolor=midblue,filecolor=midblue,urlcolor=midblue}

\definecolor{midblue}{rgb}{0,0,0.5}

\begin{document}
		
	\begin{titlepage}
		\begin{center}
			\hfill \\
			\hfill \\
			\hfill \\
			\vskip 0.5cm
			
			\title{\Large  Gravitational field of scalar lumps\\[1.5mm]
			 in higher-derivative gravity}
			
			\author{\large Luca Buoninfante
			 and Yuichi Miyashita
			 }
			
			\address{Department of Physics, Tokyo Institute of Technology, Tokyo 152-8551, Japan}
			
			\email{\rm buoninfante.l.aa@m.titech.ac.jp\\  miyashita.y.ae@m.titech.ac.jp}
			
			
			
		\end{center}

\begin{abstract}
We study the gravitational field sourced by localized scalar fields (lumps) in higher-derivative theories of gravity. By working in a static and spherically symmetric configuration, we 
find the linearized spacetime metrics generated by scalar lumps for several Lagrangians: the vanishing potential, i.e. free massive scalar field, a polynomial potential, and the tachyon potential in open string field theory.  
We perform the analysis for different theories of gravity: Einstein's general relativity, four-derivative gravity, and ghost-free nonlocal gravity. 
We discuss the limit of validity of our analysis and comment on possible future applications in the context of astrophysical compact objects.
\end{abstract}

\end{titlepage}

{
	\hypersetup{linkcolor=black}
	\tableofcontents
}

\baselineskip=17.63pt


\singlespacing

\section{Introduction}

Einstein's General Relativity (GR) has been the best theory to describe classical aspects of the gravitational interaction so far, indeed many of its predictions have been verified to a high order of precision~\cite{-C.-M.}. At the same time, there exist outstanding problems whose solution is still lacking. For instance, in the short--distance (ultraviolet) regime, GR predicts cosmological and black-hole singularities~\cite{Hawking}, whereas at the quantum level it is perturbatively non-renormalizable~\cite{tHooftVeltman74,Goroff:1985th}, thus lacking of predictability at high energies.

These open issues motivated many efforts towards formulating  a consistent quantum completion of GR. One of the most conservative approaches consists of generalizing the Einstein-Hilbert Lagrangian by including terms with fourth order derivatives of the metric tensor, such as curvatures squared like $\mathcal{R}^2$, $\mathcal{R}_{\mu\nu}\mathcal{R}^{\mu\nu}$ and $\mathcal{R}_{\mu\nu\rho\sigma}\mathcal{R}^{\mu\nu\rho\sigma}$. Very interestingly, in the seventies it was noticed that a quadratic theory of gravity is strictly renormalizable by power counting~\cite{-K.-S.}. However, it also turns out to be pathological because of the presence of higher order time-derivatives accompanied by an unhealthy  degree of freedom, i.e. a massive spin--$2$ ghost, which makes the Hamiltonian unbounded from below~\cite{Ostrogradsky:1850fid} and breaks the unitarity condition quantum mechanically (when standard prescriptions to define amplitudes are used\footnote{See, for instance, the works in Refs.~\cite{Anselmi:2017ygm,Donoghue:2019fcb,Salvio:2014soa,Salvio:2019wcp,Salvio:2019ewf,Bender:2007wu} for recent discussions on alternative mechanisms through which four-derivative theories can be made ghost-free.}).

Recently, it was realized that by considering extensions of GR according to which the Einstein-Hilbert Lagrangian is generalized through quadratic curvature terms containing \textit{nonlocal} (i.e. non-polynomial) differential operators, one can formulate a quantum theory of gravity which is also unitary~\cite{Krasnikov,Kuzmin,Moffat,Tomboulis:1997gg,Biswas:2005qr,Modesto:2011kw,Biswas:2011ar}. To avoid ghost-like degrees of freedom, the nonlocal operators are required to be specific \textit{analytic} functions of the d'Alembertian, which contain \textit{infinite order derivatives}.

In this paper we aim at computing for the first time the gravitational field generated by certain localized scalar field configurations (lumps) in the framework of higher-derivative gravity. We will work up to linear order in the metric perturbation around Minkowski background and find linearized spacetime metric solutions sourced by scalar lumps. While for the gravity sector we consider higher-derivative extensions, for the matter sector we only analyse a standard two-derivative scalar field. Furthermore, we will not make any stability analysis for the scalar lumps under investigation; for the purposes of this work it is sufficient to assume that if possible instabilities arise than the introduction of interactions with additional fields may stabilize them.

The paper is organized as follows. In Sec.~\ref{sec-action} we introduce the gravitational and scalar actions, and their expansions up to linear order around Minkowski. In Sec.~\ref{sec-free scalar} we consider a warm up exercise and compute the linearized spacetime metric for a free massive scalar field, i.e. with vanishing potential. Subsequently, in Sec.~\ref{sec-polyn scalar} and~\ref{sec-tachyon} we analyse two more realistic scalar field configurations, i.e. polynomial potential and tachyon potential in open string field theory. In both cases we compute the linearized metric and also make some discussion towards the nonlinear regime. In Sec.~\ref{conclus} we summarize our results, illustrate possible future investigations in the context of astrophysical compact objects, and draw our conclusions.

We adopt the mostly positive convention for the spacetime metric signature, i.e. $\eta=\mathrm{diag}\left(-,+,+,+\right),$ and the natural units, i.e. $c=1=\hbar.$

\section{Gravitational and scalar-field actions}\label{sec-action}

Let us consider the following action
\begin{eqnarray}
S=S_g+S_\phi\,,
\end{eqnarray}
where $S_g$ is a generalized quadratic curvature action for gravity
\begin{eqnarray}
S_g= \frac{2}{\kappa^2}\int\! {\rm d}^4x \sqrt{-g}\left[R+\frac{1}{2}\Big(RF_1(-\Box_g)R+R_{\mu\nu}F_2(-\Box_g)R^{\mu\nu}+R_{\mu\nu\rho\sigma}F_3(-\Box_g)R^{\mu\nu\rho\sigma}\Big)\right]\,,
\label{action-gravity}
\end{eqnarray}
with $\kappa^2=32\pi G$ and $G$ is Newton's constant; whereas $S_\phi$ is a generic two-derivative action for a minimally coupled real scalar field $\phi$ of mass $m$ and self-potential $V(\phi),$
\begin{eqnarray}
S_\phi=\int {\rm d}^4x \sqrt{-g}\left[\frac{1}{2}\phi (\Box_g-m^2) \phi-V(\phi)\right]\,.
\label{action-scalar}
\end{eqnarray}
The differential operators $F_i(-\Box)$ are {\it analytic} functions of the curved d'Alembertian $\Box_g=g^{\mu\nu}\nabla_{\mu}\nabla_{\nu}$, and we can expand them in Taylor series as follows
\begin{equation}
F_i(-\Box)=\sum\limits_{n=0}^N f_{i,n}(-\Box)^n\,,\qquad i=1,2,3\,,
\label{form-factor}
\end{equation}
where $f_{i,n}$ are constant coefficients, and the positive integer $N$ in general can be either finite or infinite: in the former case we have local (polynomial) theories, while in the latter nonlocal (non--polynomial) theories. It is clear that when $F_i=0$ for all $i=1,2,3$ we recover the standard two-derivative theory, i.e. Einstein's GR.

Eventually we are interested in finding the gravitational field sourced by a scalar field $\phi$ in the weak-field approximation, therefore for our purposes it is sufficient to work in a linearized regime, expanding the gravitational action up to quadratic order in the perturbation and the matter-gravity interaction up to linear order.

By perturbing the spacetime metric around Minkowski,
\begin{equation}
g_{\mu\nu}=\eta_{\mu\nu}+\kappa h_{\mu\nu}\,,\label{mink-expans}
\end{equation}
and using the relations
\begin{eqnarray}
g^{\mu\nu}&=&\eta^{\mu\nu}-\kappa h^{\mu\nu}+\kappa^2 h^{\mu}_{\rho}h^{\rho\nu}+\mathcal{O}(\kappa^3)\,,\\
\sqrt{-g}&=&\sqrt{-\eta}\left[1+\kappa\frac{h}{2}+\kappa^2\left(\frac{h^2}{8}-\frac{1}{4}h^{\rho}_{\sigma}h^{\sigma}_{\rho}\right)\right]+\mathcal{O}(\kappa^3)\,,\label{relations}
\end{eqnarray}
with $\eta\equiv {\rm det}(\eta_{\mu\nu})=-1$ and $h\equiv\eta^{\mu\nu}h_{\mu\nu},$ we can expand $S_{g}$ up to order $\mathcal{O}(\kappa^0)$ and $S_\phi$ up to order  $\mathcal{O}(\kappa);$ since we are interested in a weak-field regime we can safely neglect self-gravity interaction. 

Notice that  in this regime  one can simplify the action~\eqref{action-gravity} by neglecting the $\mathcal{F}_3(\Box_g)$-term. In fact, for analytic differential operators $F_i$, one can use the following identity
\begin{equation*}
\mathcal{R}_{\mu\nu\rho\sigma}\Box_g^n\mathcal{R}^{\mu\nu\rho\sigma}=4\mathcal{R}_{\mu\nu}\Box_g^n\mathcal{R}^{\mu\nu}-\mathcal{R}\Box_g^n\mathcal{R}+\mathcal{O}(\mathcal{R}^3)+{\rm div}\,,
\end{equation*}
where {\rm div} stands for total derivative terms; thus up to order $\mathcal{O}(h^2)$ the Riemann--squared contribution can be replaced by combinations of Ricci scalar and Ricci tensor squared plus boundary terms and higher order contributions $\mathcal{O}(\mathcal{R}^3)\sim \mathcal{O}(h^3)$ which are negligible in our regime of interest. Therefore, in the rest of the discussion we can safely neglect the Riemann contribution in $S_g;$ for simplicity we keep the same notation for $F_1$ and $F_2,$ and set $F_3=0$.

Hence, the relevant terms up to order $\mathcal{O}(h^2)$ in the gravitational action are given by\footnote{See also Refs.~\cite{Biswas:2016etb,Koshelev:2017tvv,SravanKumar:2019eqt} for studies of metric perturbations in higher-derivative gravity around maximally symmetric and more general backgrounds.}~\cite{Biswas:2011ar}:
\begin{equation}
\begin{array}{rl}
S_g=&\displaystyle\!\! \int {\rm d}^4x\left\lbrace \frac{1}{2}h_{\mu\nu}f(\Box)\Box h^{\mu\nu}-h_{\mu}^{\sigma}f(\Box)\partial_{\sigma}\partial_{\nu}h^{\mu\nu}+hg(\Box)\partial_{\mu}\partial_{\nu}h^{\mu\nu}\right.\\[3mm]
& \,\,\,\,\,\,\,\,\,\,\,\,\,\,\,\,\,\,\,\,\,\,\,\,\,\,\,\,\,\,\displaystyle \left. -\frac{1}{2}hg(\Box)\Box h +\frac{1}{2}h^{\lambda\sigma}\frac{f(\Box)-g(\Box)}{\Box}\partial_{\lambda}\partial_{\sigma}\partial_{\mu}\partial_{\nu}h^{\mu\nu}\right\rbrace \,,
\label{lin-quad-action}
\end{array}
\end{equation}
where 
\begin{equation}
\begin{array}{rl}
f(\Box)\equiv & \displaystyle  1+\frac{1}{2}F_2(\Box)\Box,\\
g(\Box)\equiv & \displaystyle 1-2F_1(\Box)\Box-\frac{1}{2}F_2(\Box)\Box\,.
\label{f-g}
\end{array}
\end{equation}
Note that $\Box=\eta^{\mu\nu}\partial_\mu\partial_\nu$ (without subscript `$g$') stands for the flat d'Alembertian.

Whereas, the scalar field action up to linear order in the metric perturbation reads
\begin{eqnarray}
S_\phi = \frac{1}{2}\int {\rm d}^4x \,\phi (\Box-m^2)\phi + \frac{1}{2}\kappa\int {\rm d}^4x \,h^{\mu\nu}T_{\mu\nu}\,,\label{nonlocal-scalar1}
\end{eqnarray}
where $T_{\mu\nu}$ is the stress-energy of the scalar field $\phi$ defined as
\begin{eqnarray}
T_{\mu\nu}=-\frac{2}{\sqrt{-g}}\frac{\delta S_\phi}{\delta g^{\mu\nu}}\simeq \frac{2}{\kappa}\frac{\delta S_\phi}{\delta h^{\mu\nu}}\,,\label{stress-def}
\end{eqnarray}
and one can easily show that
\begin{eqnarray}
T_{\mu\nu}= \partial_\mu\phi\,\partial_\nu\phi-\frac{1}{2}\eta_{\mu\nu}\partial_\rho\phi\,\partial^\rho\phi-\frac{1}{2}\eta_{\mu\nu}m^2\phi^2-\eta_{\mu\nu}V(\phi)\,.\label{local-stress}
\end{eqnarray}

\subsection{Linearized field equations and gravitational potentials}\label{sec-lin-eq}

We now wish to obtain the field equations for the metric perturbation $h_{\mu\nu}$ and for the scalar field $\phi$ in the linear regime. We work in a \textit{static} and \textit{spherically symmetric} configuration.

By varying the action in Eq.~\eqref{lin-quad-action} with respect to $h_{\mu\nu},$ we obtain the linearized field equations for the metric perturbation~\cite{Biswas:2011ar,Biswas:2013cha}
\begin{equation}
\begin{array}{ll}
\displaystyle f(\Box)\left(\Box h_{\mu\nu}-\partial_{\sigma}\partial_{\nu}h_{\mu}^{\sigma}-\partial_{\sigma}\partial_{\mu}h_{\nu}^{\sigma}\right) \displaystyle +g(\Box)\left(\eta_{\mu\nu}\partial_{\rho}\partial_{\sigma}h^{\rho\sigma}+\partial_{\mu}\partial_{\nu}h-\eta_{\mu\nu}\Box h\right)&\\[3mm]
\,\,\,\,\,\,\,\,\,\,\,\,\,\,\,\,\,\,\,\,\,\,\,\,\,\,\,\,\,\,\,\,\,\,\,\,\,\,\,\,\,\,\,\,\,\,\,\,\,\,\,\,\,\,\,\,\,\,\,\,\,\,\,\,\,\,\,\,\,\,+\displaystyle \frac{f(\Box)-g(\Box)}{\Box}\partial_{\mu}\partial_{\nu}\partial_{\rho}\partial_{\sigma}h^{\rho\sigma}=-\frac{\kappa}{2} T_{\mu\nu}\,.&
\label{lin-field-eq}
\end{array}
\end{equation}
In what follows we assume that the gravitational action $S_g$ only propagates a spin-$2$ degree of freedom on-shell; this corresponds to choose~\cite{Biswas:2011ar}
\begin{equation}
f(\Box)=g(\Box)\quad \Leftrightarrow\quad 2F_1(\Box)=-F_2(\Box)\,,\label{form-factos choice}
\end{equation}
which is already a very good model for the purposes of this paper. In other words, we only consider higher-derivative generalizations of the massless transverse spin-$2$ degree of freedom without introducing any scalaron\footnote{Indeed, the gauge independent part of the graviton propagator around Minkowski is given by~\cite{Tomboulis:1997gg,Biswas:2011ar}
	\begin{equation} \label{Propagator}
	\Pi_{\mu\nu\rho\sigma}(k)=\frac{1}{f}\Pi^{\rm GR}_{\mu\nu\rho\sigma}(k)+\frac{3}{2}\frac{(f-g)}{f(f-3g)k^2}\mathcal{P}^{0}_{s,\,\mu\nu\rho\sigma}\,,
	\end{equation}
	where $f\equiv f(-k^2),$ $g\equiv g(-k^2)$ in momentum space, and $\mathcal{P}^2_{\mu\nu\rho\sigma}$ and $\mathcal{P}^{0}_{s,\,\mu\nu\rho\sigma}$ are the so called spin-$2$ (traceless and transverse) and spin-$0$ (trace) projector operators~\cite{Barnes-Rivers,VanNieuwenhuizen:1973fi}. It is now clear that, when the condition $f=g$ is satisfied, then the propagating scalar contribution disappears and we are only left with the massless and transverse spin-$2$ as in Einstein's GR.}.

By working in the harmonic gauge, $\partial^\mu (h_{\mu\nu}-\eta_{\mu\nu}h/2)=0,$ the field equations~\eqref{lin-field-eq} become
\begin{equation}
f(\Box)\Box\left(h_{\mu\nu}-\frac{1}{2}\eta_{\mu\nu}h\right)=-\frac{\kappa}{2}T_{\mu\nu}\,.
\label{lin-field-eq-harmonic}
\end{equation}
Then, we can consistently impose the following metric ansatz written in isotropic coordinates
\begin{equation}
{\rm d}s^2=-(1+2\Phi){\rm d}t^2 +(1-2\Psi)({\rm d}r^2+r^2{\rm d}\Omega^2)\,,\label{isotr-metric}
\end{equation}
where $r=\sqrt{x^2+y^2+z^2}$ and ${\rm d}r^2+r^2{\rm d}\Omega^2={\rm d}x^2+{\rm d}y^2+{\rm d}z^2,$ while $\kappa h_{00}=-2\Phi$ and $\kappa h_{ij}=-2\Psi\delta_{ij}$ are the metric potentials sourced by $T_{\mu\nu}.$ By making use of the staticity and spherical symmetry assumptions one can show that the metric potentials in Eq.~\eqref{isotr-metric} are the solutions of the following generalized Poisson equations
\begin{eqnarray}
&&\displaystyle f(\nabla^2)\nabla^2\Phi(r)=4\pi G\,\left\lbrace T[\phi](r)+2T_{00}[\phi](r)\right\rbrace\,,\label{field-eq-pot-phi}\\[2mm]
&&\displaystyle f(\nabla^2)\nabla^2\Psi(r)=4\pi G \,T_{00}[\phi](r)\,;
\label{field-eq-pot-psi}
\end{eqnarray}
$f(\nabla^2)$ is now a function of the spatial Laplacian $\nabla^2=\delta^{ij}\partial_i\partial_j$ as we are working in a static configuration.
When $f=g=1,$ which also means $F_1=F_2=0,$ the differential equations in Eqs.~\eqref{field-eq-pot-phi} and~\eqref{field-eq-pot-psi} reduce to the GR case.

Hence, we have now established the necessary framework to compute the linearized spacetime metric sourced by a minimally coupled scalar field. We have to solve the scalar field equation
\begin{equation}
\left(\nabla^2-m^2\right)\phi(x)=\frac{{\rm d }V(\phi)}{{\rm d}\phi}\,,
\label{lin-field-scalar}
\end{equation}
and substitute its solution into the components of $T_{\mu\nu}[\phi].$ Then, we can solve either analytically or numerically Eqs.~(\ref{field-eq-pot-phi},\ref{field-eq-pot-psi}) to find $\Phi$ and $\Psi.$ Note that in Eq.~\eqref{lin-field-scalar} we consistently neglected higher-order terms $\mathcal{O}(\kappa)$ in order to obtain the metric perturbation $\Phi$ and $\Psi$ up to $\mathcal{O}(\kappa^2\sim G)$ from Eq.~\eqref{lin-field-eq-harmonic}.

In what follows we study three type of scalar field potentials: vanishing potential, i.e. free massive scalar field, in Sec.~\ref{sec-free scalar} as a warm up exercise; a polynomial potential in Sec.~\ref{sec-polyn scalar}; and the tachyon potential in open string field theory in Sec.~\ref{sec-tachyon}. For all the cases we perform the analysis in three different theories of gravity:\textit{ (i)} Einstein's GR; \textit{(ii)} four-derivative gravity; \textit{(iii)} ghost-free nonlocal gravity. Such a distinction will allow us to understand the role of nonlocality in both gravity and matter sectors.

To be more specific, we will consider the following form factors:
\begin{eqnarray}
&&\text{\textit{(i)} Einstein's GR:}\quad 2F_1=F_2=0\,\, \Rightarrow\,\, f(\Box)=1\,,\\[2mm]
&&\text{\textit{(ii)} Four-derivative gravity:}\quad 2F_1=F_2=-\alpha\,\, \Rightarrow\,\, f(\Box)=1-\frac{\alpha}{2}\Box\,,\\
&&\text{\textit{(iii)} Ghost-free nonlocal gravity:}\quad 2F_1=F_2=2\frac{e^{-\Box/\mu^2}-1}{\Box}\,\, \Rightarrow\,\,f(\Box)=e^{-\Box/\mu^2}\,,\qquad
\end{eqnarray}
where $\alpha>0$ is a constant and it is physically related to the mass of the spin-$2$ massive ghost in four-derivative gravity, and $\mu$ is the energy scale of nonlocality at which new gravitational physics is expected to manifest. The chosen nonpolynomial form factor for the third model gives a ghost-free propagator around Minkowski background, so that the full gravity theory turns out to be unitary despite the presence of higher (infinite) order time-derivatives~\cite{Tomboulis:2015gfa,sen-epsilon,carone,Briscese:2018oyx,chin,Buoninfante:2018mre,Koshelev:2021orf}. 

Note that, especially in the recent years, lots of studies on linearized spacetime metric solutions have been made in the context of both four-derivative and infinite-derivative gravity~\cite{Tseytlin:1995uq,Siegel:2003vt,Edholm:2016hbt,Frolov:2015bia,Frolov,Frolov:2015usa,Buoninfante:2018xiw,Koshelev:2018hpt,Buoninfante:2018rlq,Buoninfante:2018stt,Buoninfante:2018xif,Boos:2018bxf,Kilicarslan:2018yxd,Buoninfante:2019swn,Buoninfante:2018lnh,Buoninfante:2020ctr,Boos:2020kgj,Abel:2019zou,Giacchini:2018wlf,Burzilla:2020bkx,Burzilla:2020utr,Boos:2020ccj,Kolar:2020bpo,Boos:2021suz,Boos:2018bhd}, see also Refs.~\cite{Kilicarslan:2019njc,Dengiz:2020xbu,Kolar:2021rfl} for some exact solutions. Most of these studies only concerned with regularization of the Newtonian potential in presence of singular point-like sources. In this work we compute for the first time the gravitational metric potentials in higher-derivative theories of gravity in presence of scalar field lumps. We will see that, even when the source is already regular, higher (especially infinite) derivative can still play a crucial role by modifying the short-distance behavior.

\section{Free massive scalar field}\label{sec-free scalar}

In this section, as a warm up exercise we consider a free massive scalar field, i.e. characterized by a vanishing potential $V(\phi)=0.$ The scalar field equation reads
\begin{equation}
\left(\nabla^2-m^2\right)\phi(r)=0\,,
\label{lin-field-scalar-free}
\end{equation}
and consistently with the assumptions of  staticity and spherical symmetry, i.e. $\phi\equiv \phi(r),$ we get
\begin{equation}
\Big(\partial_r^2+\frac{2}{r}\partial_r-m^2\Big)\phi(r)=0\quad \Rightarrow\quad \phi(r)=C \frac{e^{-mr}}{r}+D \frac{e^{mr}}{r}\,,
\label{sol-scalar}
\end{equation}
where $C,$ $D$ are two integration constants. Since we are interested in a localized field configuration and eventually in asymptotically flat metrics, we set $D=0$. Thus, the scalar field solution
\begin{equation}
\phi(r)=C\frac{e^{-mr}}{r}\,\label{sol-scalar2}
\end{equation}
could be regarded as a localized singular source of radius $\sim 1/m;$
the constant $C$ will enter in the redefinition of Newton's constant as we will see below.


The stress-energy tensor is given by the expression in Eq.~\eqref{local-stress}, and for a free scalar field it reduces to
\begin{equation}
T_{\mu\nu}=\partial_\mu\phi\,\partial_\nu\phi-\frac{1}{2}\eta_{\mu\nu}\partial_\rho\phi\,\partial^\rho\phi-\frac{1}{2}\eta_{\mu\nu}m^2\phi^2\,,
\end{equation}
whose $(00)$-component and trace read
\begin{eqnarray}
T_{00}= C^2 \frac{1+2mr+2m^2r^2}{2r^4}e^{-2mr}\,,
\end{eqnarray}
and
\begin{eqnarray}
T= -C^2 \frac{1+2mr+3m^2r^2}{2r^4}e^{-2mr}\,.
\end{eqnarray}

\subsection{General relativity}

Given the above expressions for the stress-energy tensor components, in the GR case ($f=1$) the field equations for the metric potentials read
\begin{eqnarray}
&&\nabla^2\Phi(r)=-4\pi G\,C^2 m^2\frac{e^{-2mr}}{r^2}\,,\nonumber\\[2mm]
&& \nabla^2\Psi(r)=4\pi G\,C^2\frac{1+2mr+2m^2r^2}{2r^4}e^{-2mr}\,,
\label{field-eq-pot-free-local}
\end{eqnarray}
and can be solved analytically in terms of homogeneous and particular solutions:
\begin{eqnarray}
&& \Phi(r)=  -\frac{A_1}{r}-2\pi GmC^2\frac{e^{-2mr}}{r}-4\pi GC^2m^2 {\rm Ei}\left(-2mr\right)+A_2\,,\\[2mm]
&& \Psi(r)=-\frac{B_1}{r}+\pi GC^2\frac{e^{-2mr}}{r^2}+B_2\,,
\end{eqnarray}
where ${\rm Ei}(-x)=-\int_x^{\infty}{\rm d}t\,\frac{e^{-t}}{t}$ is the so called exponential integral function, and $A_1,$ $A_2$, $B_1,$ $B_2$ are four integration constants. Since we are interested in asymptotically flat solutions we set $A_2=0=B_2.$ It is clear that if $A_1,B_1>0,$ then $-A_1/r$ and $-B_1/r$ are the standard Newtonian components which dominate at large distances, i.e. for $r\gg 1/m$.

Hence, by also rescaling  Newton's constant as $G\rightarrow \bar{G}=2\pi GC^2,$ the gravitational potentials sourced by a free massive scalar field in Einstein's GR are
\begin{eqnarray}
&& \Phi(r)=   -\frac{A_1}{r}-\bar{G}m\frac{e^{-2mr}}{r}-2 \bar{G}m^2 {\rm Ei}\left(-2mr\right)\,,\nonumber\\[2mm]
&& \Psi(r)=  -\frac{B_1}{r}+\bar{G}\frac{e^{-2mr}}{2r^2}\,.
\label{grav-sol-Local-local}
\end{eqnarray}
We can notice that $\Phi$ is always negative while $\Psi$ becomes positive at short distances $r\lesssim 1/m$, which means that the relativistic component $\Psi$ beyond the Newtonian approximation introduces a repulsive contribution. More explicitly, the short-distance behavior of the two potentials is given by $\Phi\sim -(A_1+\bar{G}m)/r$ and $\Psi\sim \bar{G}/(2r^2)$, and both show a singularity. The behavior of the metric potentials  is shown in Fig.~\ref{fig1}.

\subsection{Four-derivative gravity}

In the case of four-derivative gravity the field equations for the metric potentials are
\begin{eqnarray}
&&\left(1-\frac{\alpha}{2}\nabla^2\right)\nabla^2\Phi(r)=-4\pi G\,C^2 m^2\frac{e^{-2mr}}{r^2}\,,\nonumber\\[2mm]
&&\left(1-\frac{\alpha}{2}\nabla^2\right) \nabla^2\Psi(r)=4\pi G\,C^2\frac{1+2mr+2m^2r^2}{2r^4}e^{-2mr}\,.
\label{field-eq-pot-free-four}
\end{eqnarray}
These equations can not be solved analytically but we can write them in a more suitable form which allows us to solve them at least numerically. Indeed, by using the following Fourier transforms
\begin{eqnarray}
\int {\rm d}^3r\, \frac{e^{-2mr}}{r}e^{-i\vec{k}\cdot\vec{r}}&=& \frac{4\pi}{k^2+4m^2} \,,\\[2mm]
\int {\rm d}^3r\, \frac{e^{-2mr}}{r^2}e^{-i\vec{k}\cdot\vec{r}}&=& \frac{4\pi \,{\rm arccot}\left(2m/k\right)}{k}  \,,\\[2mm]
\int {\rm d}^3r\, {\rm Ei}(-2mr)e^{-i\vec{k}\cdot\vec{r}}&= & \frac{4\pi}{k^3}\left[\frac{2km}{k^2+4m^2}-{\rm arccot}\left(\frac{2m}{k}\right)  \right]\,,\,\,\,\,\,
\label{fourier transfor-partic}
\end{eqnarray}
we can recast the metric potentials in terms of the anti-transform integrals:
\begin{eqnarray}
\Phi(r)&=& -\frac{A_1}{r}-\frac{2\bar{G}m}{\pi r}\int_0^{\infty} {\rm d}k\,\left[1-\frac{2m}{k}\,{\rm arccot}\left(\frac{2m}{k}\right)   \right]\frac{\sin(kr)}{k(1+\alpha k^2/2)}\,,\\[2mm]
\Psi(r)&=&  -\frac{B_1}{r}+\frac{\bar{G}}{\pi r}\int_0^{\infty}  {\rm d}k\,{\rm arccot}\left(\frac{2m}{k}\right) \,\frac{\sin(kr)}{(1+\alpha k^2/2)} \,.
\label{pot-nonlocal-grav-free-four}
\end{eqnarray}
As extra boundary conditions we set to zero the Yukawa contributions in the homogeneous solution as they are not important for our purposes. Thus, the only homogeneous term is the standard Newtonian one. The numerical solutions of the integrals in Eqs.~\eqref{pot-nonlocal-grav-free-four} are shown in Fig.~\ref{fig1}.

\subsection{Ghost-free nonlocal gravity}

In the case of nonlocal gravity the field equations for the metric potentials are
\begin{eqnarray}
&&e^{-\nabla^2/\mu^2}\nabla^2\Phi(r)=-4\pi G\,C^2 m^2\frac{e^{-2mr}}{r^2}\,,\nonumber\\[2mm]
&&e^{-\nabla^2/\mu^2} \nabla^2\Psi(r)=4\pi G\,C^2\frac{1+2mr+2m^2r^2}{2r^4}e^{-2mr}\,.
\label{field-eq-pot-free-loc-nonloc}
\end{eqnarray}
%

\begin{figure}[t!]
	\centering
	\subfloat[Subfigure 1 list of figures text][]{
		\includegraphics[scale=0.445]{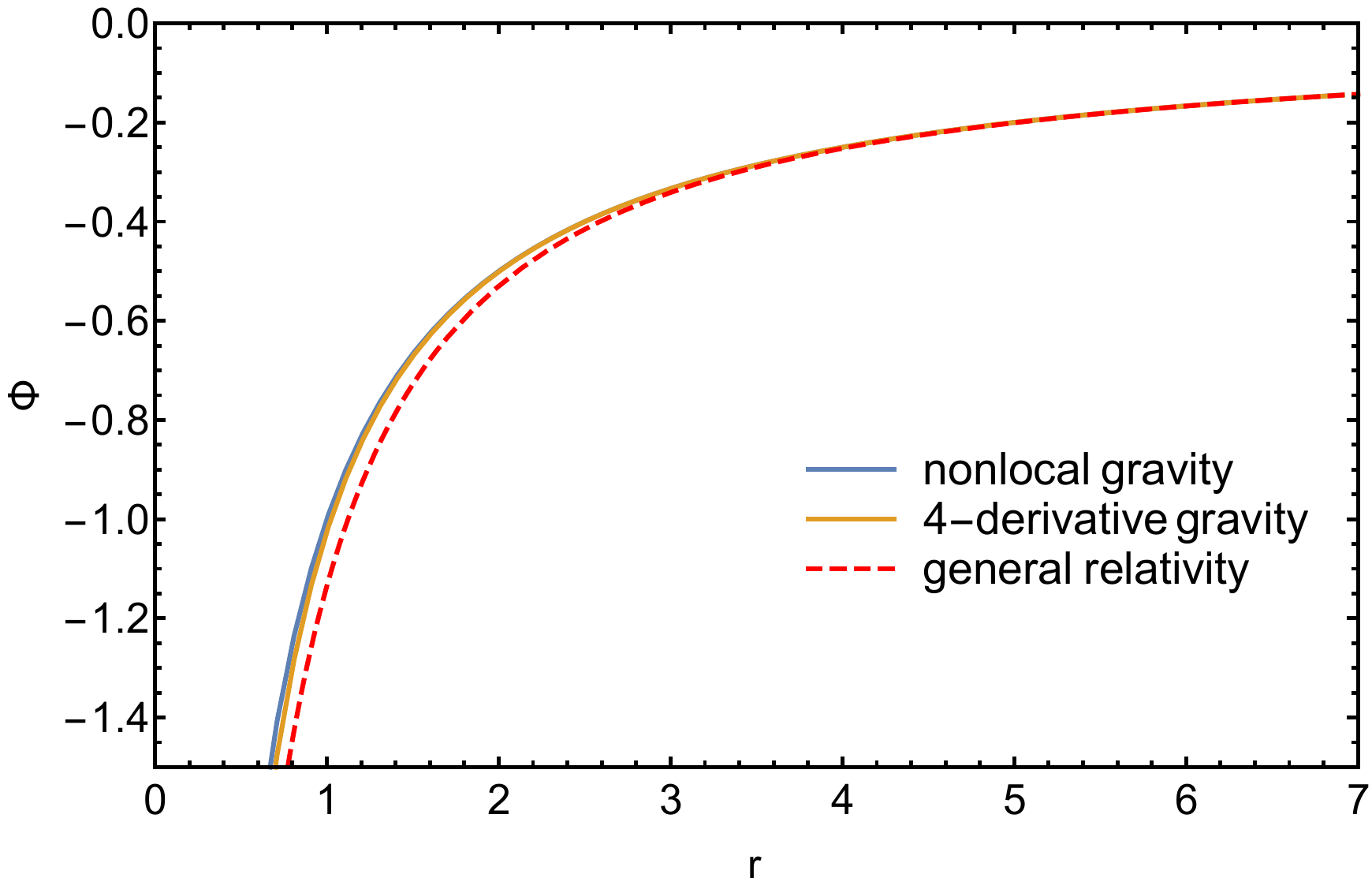}}
	\subfloat[Subfigure 2 list of figures text][]{
		\includegraphics[scale=0.445]{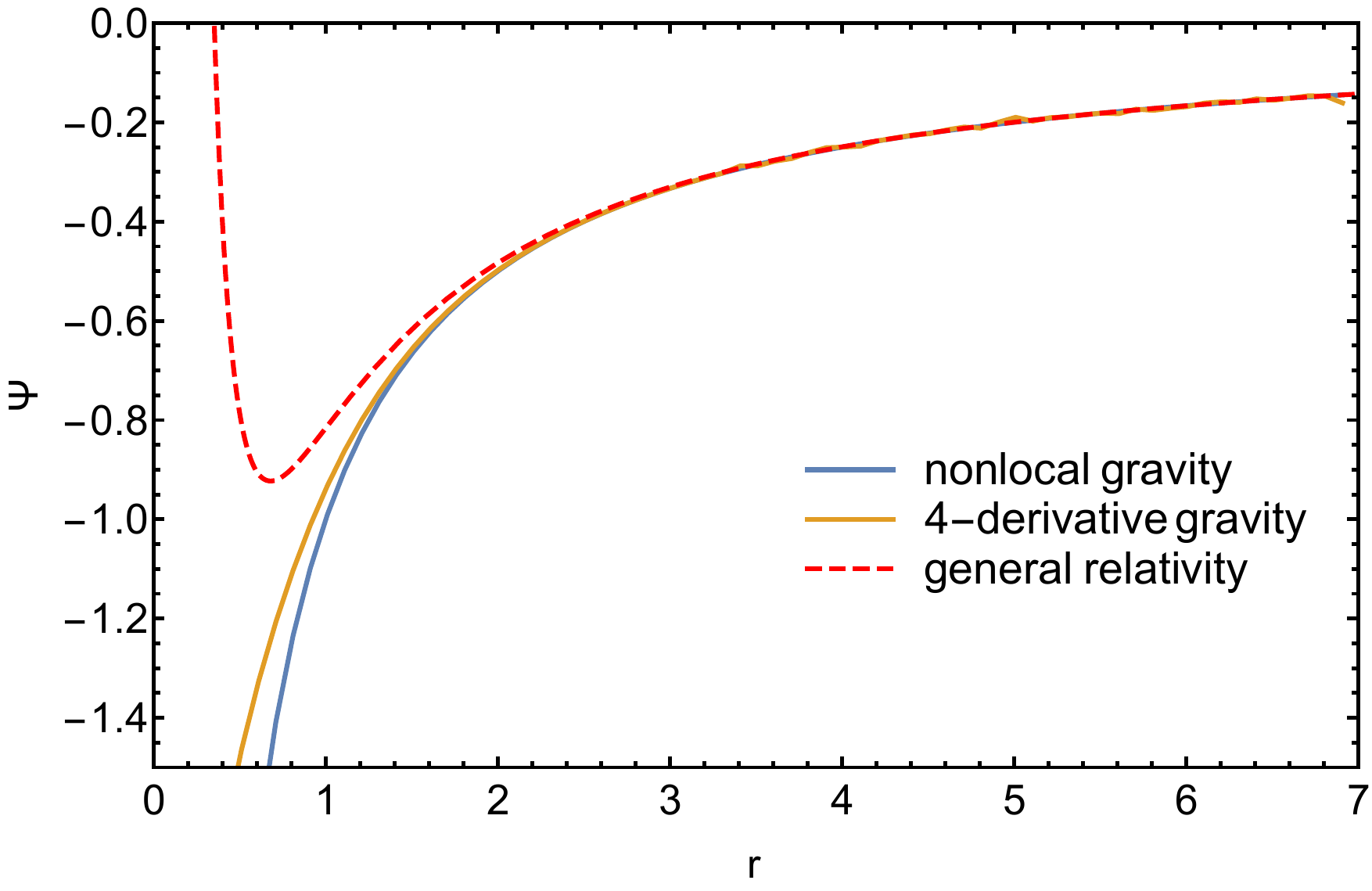}
	}
	\protect\caption{Behaviors of the metric potentials (a) $\Phi$ and (b) $\Psi$ sourced by a free massive scalar field. The blue solid line corresponds to the nonlocal gravity case with $f(\nabla^2)=e^{-\nabla^2/\mu^2};$ the orange solid line to the four-derivative case with $f(\nabla^2)=1-\alpha\nabla^2/2;$ whereas the red dashed line to the GR case $f(\nabla^2)=1.$ We set $A_1=B_1=1,$ $G=1/2,$ $m=1$ and $\mu=0.1=\alpha.$ 
	}\label{fig1}
\end{figure}

If we introduce the field redefinitions  $\tilde{\Phi}=e^{-\nabla^2/\mu^2}\Phi$ and $\tilde{\Psi}=e^{-\nabla^2/\mu^2}\Psi,$ the asymptotically flat solutions for $\tilde{\Phi}$ and $\tilde{\Psi}$ turn out to be equal to the ones obtained in Eq.~\eqref{grav-sol-Local-local} in GR. Then, by inverting back to the original metric potentials we have
\begin{eqnarray}
&& \Phi(r)=   -A_1e^{\nabla^2/\mu^2}\left(\frac{1}{r}\right)-\bar{G}me^{\nabla^2/\mu^2}\left(\frac{e^{-2mr}}{r}\right)-2 \bar{G}m^2 e^{\nabla^2/\mu^2} \Big({\rm Ei}(-2mr)\Big)\,,\\[2mm]
&& \Psi(r)=  -B_1e^{\nabla^2/\mu^2}\left(\frac{1}{r}\right)+\frac{\bar{G}}{2}e^{\nabla^2/\mu^2}\left(\frac{e^{-2mr}}{r^2}\right)\,.
\label{pot-nonlocal-acting}
\end{eqnarray}
First of all, we can notice that for the homogeneous parts we have  $\nabla^2(1/r)=0$ which implies $e^{\nabla^2/\mu^2}(1/r)=1/r,$ consistently with the fact that nonlocality does not affect the homogeneous solutions~\cite{Barnaby:2007ve}. 
As for the particular solutions, the action of the operator $e^{\nabla^2/\mu^2}$ is non-trivial and can be evaluated at least numerically by means of Fourier-transform method. Indeed, by using the Fourier transforms~\eqref{fourier transfor-partic}
we can recast the metric potentials in terms of the anti-transform integrals:
\begin{eqnarray}
 \Phi(r)&=& -\frac{A_1}{r}-\frac{2\bar{G}m}{\pi r}\int_0^{\infty} {\rm d}k\,e^{-k^2/\mu^2}\left[1-\frac{2m}{k}\,{\rm arccot}\left(\frac{2m}{k}\right)   \right]\frac{\sin(kr)}{k}\,,\\[2mm]
\Psi(r)&=&  -\frac{B_1}{r}+\frac{\bar{G}}{\pi r}\int_0^{\infty}  {\rm d}k\,e^{-k^2/\mu^2}\,{\rm arccot}\left(\frac{2m}{k}\right) \,\sin(kr) \,.
\label{pot-nonlocal-grav-free}
\end{eqnarray}
We have computed the above integrals numerically and shown the behaviors of the solutions in Fig.~\ref{fig1} also in comparison with the four-derivative and GR cases. 

It is clear that the divergence in the limit $r\rightarrow 0$ is still present as $A_1,B_1\neq 0.$ If we would impose the boundary conditions $A_1=0=B_1,$ then the two potentials would turn out to be fully regularized by nonlocality but at the same time they would not recover the $1/r$ fall-off at large distances.

Let us now make the following remark regarding the singularity at $r=0.$ It is often said that in nonlocal theories the Newtonian potential is regularized. However, what is usually considered is a differential equation with a Dirac delta source on the right-hand-side, e.g. $e^{-\nabla^2/\mu^2}\nabla^2\Phi=4\pi G m\delta^{3}(\vec{r}).$ In such a case, by solving for $\tilde{\Phi}$ one finds two Newtonian contributions: one coming from the homogenous part and the other from the particular part, such that $\nabla^2 (1/r)_{\rm hom}=0$ and $\nabla^2(1/r)_{\rm part}\neq 0.$  Then, one sets the homogeneous solution equal to zero as a boundary condition, so that the particular one is the only left, and it is indeed the solution that can be modified and ameliorated by nonlocality. 


In Sec.~\ref{sec-polyn scalar} and~\ref{sec-tachyon} we will see that for more realistic scenarios of non-vanishing potential, e.g. polynomial scalar potential and the tachyon potential in open string field theory, the scalar lump solution sources a gravitational field that recovers the Newtonian behavior at large distances even when the homogeneous part is set equal to zero. This feature, indeed, will allow us to find sensible nonsingular solutions which recover the standard $1/r$-law at large distances.

\section{Polynomial scalar field potential}\label{sec-polyn scalar}

In this section we consider a self-interacting massless  ($m=0$) scalar field characterized by the following polynomial potential
\begin{eqnarray}
V(\phi)=\frac{g}{3}\phi^3-\frac{\lambda}{4}\phi^4\,,\label{polyn-pot}
\end{eqnarray}
where $g$ and $\lambda$ are two positive coupling constants.
The scalar field equation in Eq.~\eqref{lin-field-eq} reads
\begin{eqnarray}
\left(\partial_r^2+\frac{2}{r}\partial_r\right) \phi(r)=g \phi^2(r)-\lambda\phi^3(r)\,,
\end{eqnarray}
and it admits the following exact solution
\begin{eqnarray}
\phi(r)=\frac{2}{g}\frac{1}{r^2+R^2}\,,\qquad R=\frac{1}{g}\sqrt{\frac{\lambda}{2}}\,,\label{lump-polyn}
\end{eqnarray}
which describes a localized scalar field configuration, i.e. a lump, and the parameter $R$ can be interpreted as the radius of the system. 

The components of the stress-energy tensor necessary for the computation of the linearized spacetime metric are given by
\begin{eqnarray}
T_{00}(r)&=&\frac{16}{3g^2}\frac{2r^2-R^2}{(r^2+R^2)^4}\,,\\
T(r)+2T_{00}(r)&=& -\frac{16}{3g^2}\frac{r^2-2R^2}{(r^2+R^2)^4}\,.
\end{eqnarray}
The mass-energy $M$ of such a lump configuration is
\begin{eqnarray}
M=\int {\rm d}^3r\,T_{00}(r)=\frac{2\pi^2}{3g^2R^3}\,.\label{mass-polyn}
\end{eqnarray}
We can also define the corresponding Schwarzschild radius
\begin{eqnarray}
R_{\rm sch}=2GM=\frac{4\pi^2G}{3g^2R^3}\,.
\end{eqnarray}
Therefore, our weak-field approximation remains valid as long $R\gtrsim R_{\rm sch},$ which imposes the following relation among the coupling constants $g$ and $\lambda$ in the scalar potential~\eqref{polyn-pot}:
\begin{eqnarray}
\lambda\gtrsim \left(4\pi\sqrt{G/3}\right)g\,.\label{ineq-polyn}
\end{eqnarray}
Violating the above inequality would imply that we are in a regime in which black-hole formation can happen and we can not trust the weak-field approximation.

Let us now compute the linearized spacetime metric potentials for the three different theories of gravity as done for the free massive scalar field in the previous section.

\subsection{General relativity}

In the GR case ($f=1$) the field equations for the metric potentials are
\begin{eqnarray}
&&\nabla^2\Phi(r)= - G\,\frac{64\pi}{3g^2}\frac{r^2-2R^2}{(r^2+R^2)^4}\,,\nonumber\\[2mm]
&& \nabla^2\Psi(r)=G\,\frac{64\pi}{3g^2}\frac{2r^2-R^2}{(r^2+R^2)^4}\,.
\label{field-eq-pot-polyn}
\end{eqnarray}
By using the following Fourier transforms:
\begin{eqnarray}
\int {\rm d}^3r\,  \frac{r^2-2R^2}{(r^2+R^2)^4}e^{-i\vec{k}\cdot\vec{r}}&=& -\frac{\pi^2}{8R^3}e^{-Rk}\left(1+kR+k^2R^2\right)  \,,\nonumber\\[2mm]
\int {\rm d}^3r\, \frac{2r^2-R^2}{(r^2+R^2)^4}e^{-i\vec{k}\cdot\vec{r}}&=& \frac{\pi^2}{8R^3}e^{-Rk}\left(1+kR-k^2R^2\right)  \,,
\label{fourier transfor-polyn}
\end{eqnarray}
we can write the solutions as follows
\begin{eqnarray}
&& \Phi(r)=  -\frac{A_1}{r}-\frac{4\pi G}{3g^2}\left(\frac{2}{(r^2+R^2)^2}+\frac{1}{R^2}\frac{1}{r^2+R^2}+\frac{1}{R^3r}\arctan\left(\frac{r}{R}\right)\right)+A_2\,,\\[2mm]
&& \Psi(r)=-\frac{B_1}{r}+\frac{4\pi G}{3g^2}\left(\frac{2}{(r^2+R^2)^2}-\frac{1}{R^2}\frac{1}{r^2+R^2}-\frac{1}{R^3r}\arctan\left(\frac{r}{R}\right)\right)+B_2\,.
\end{eqnarray}
Note that now the large-distance behavior of the particular solutions for both $\Phi$ and $\Psi$ is given by $-(2G\pi^2/3g^2R^3)\frac{1}{r}=-GM\frac{1}{r},$ i.e. by a Newtonian fall-off. Thus, we can safely set $A_1=0=B_1$ and still recover Newton's law far from the source. We also impose the usual asymptotic flatness condition  $A_2=0=B_2.$ 

We can rewrite the metric potentials in terms of the mass-energy $M$ of the system introduced in Eq.~\eqref{mass-polyn} as follows
\begin{eqnarray}
&& \Phi(r)= -\frac{2GM}{\pi}\left(\frac{2R^3}{(r^2+R^2)^2}+\frac{R}{r^2+R^2}+\frac{1}{r}\arctan\left(\frac{r}{R}\right)\right)\,,\\[2mm]
&& \Psi(r)=\frac{2GM}{\pi}\left(\frac{2R^3}{(r^2+R^2)^2}-\frac{R}{r^2+R^2}-\frac{1}{r}\arctan\left(\frac{r}{R}\right)\right)\,,
\end{eqnarray}
so that the large-distance limit is now given by the standard Newtonian expressions
\begin{eqnarray}
r\gg R \quad \Rightarrow \quad \Phi,\Psi \sim -\frac{GM}{r}\,.
\end{eqnarray}
Whereas, in the short-distance regime we have
\begin{eqnarray}
r\rightarrow 0 \quad \Rightarrow \quad \Phi \rightarrow -\frac{8GM}{\pi R},\quad \Psi\rightarrow 0\,.
\end{eqnarray}
From the last expression, we understand that the linear regime can be trusted as long as $2|\Phi|,\,2|\Psi|<1,$ i.e. $R>16GM/\pi ,$ which is consistent with Eq.~\eqref{ineq-polyn}, namely with the fact that lump's radius must be larger than the corresponding Schwarzschild radius.

The behavior of the metric potentials is shown in Fig.~\ref{fig2}.

\subsection{Four-derivative gravity}

\begin{figure}[t!]
	\centering
	\subfloat[Subfigure 1 list of figures text][]{
		\includegraphics[scale=0.428]{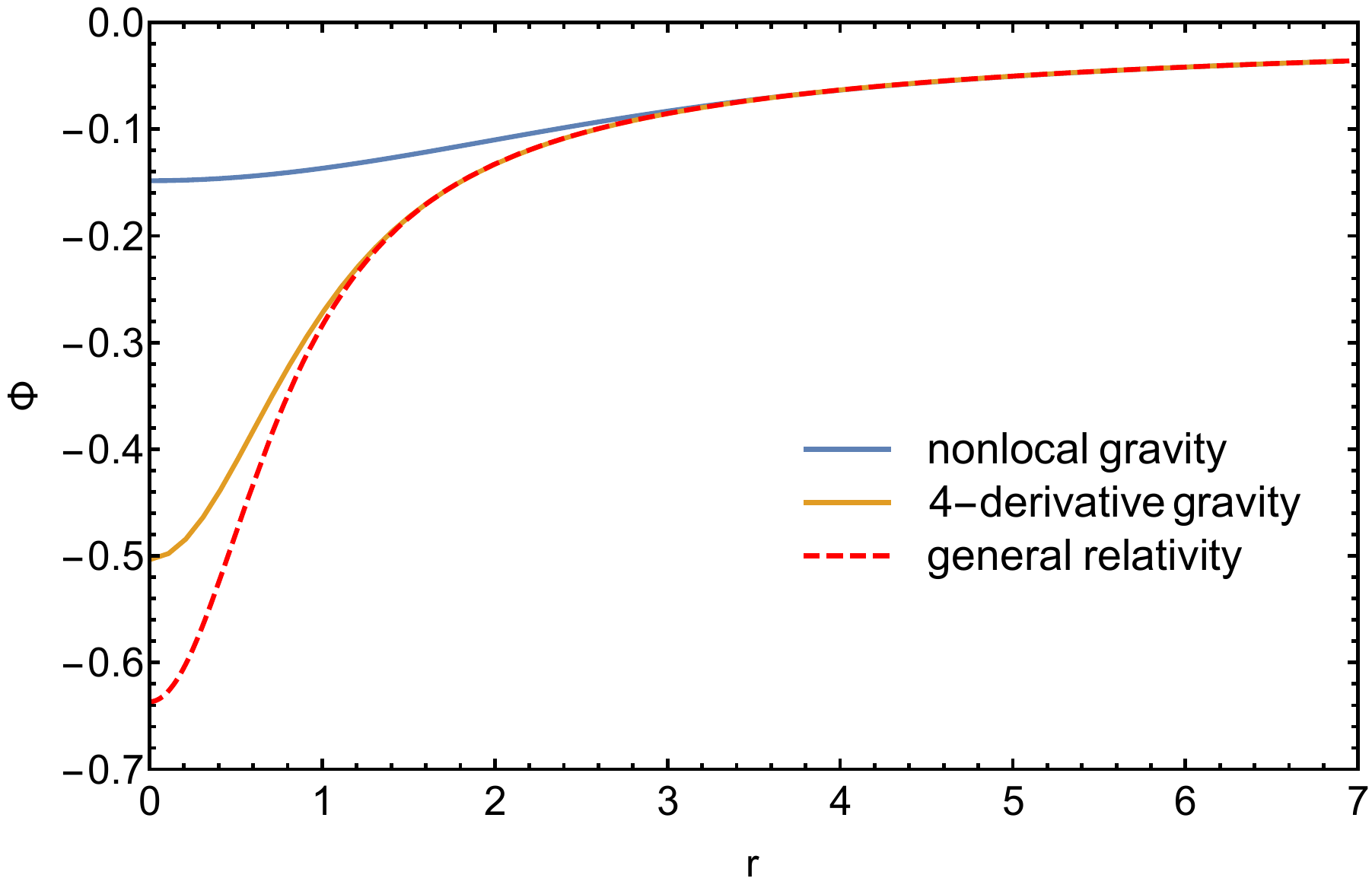}}
	\subfloat[Subfigure 2 list of figures text][]{
		\includegraphics[scale=0.428]{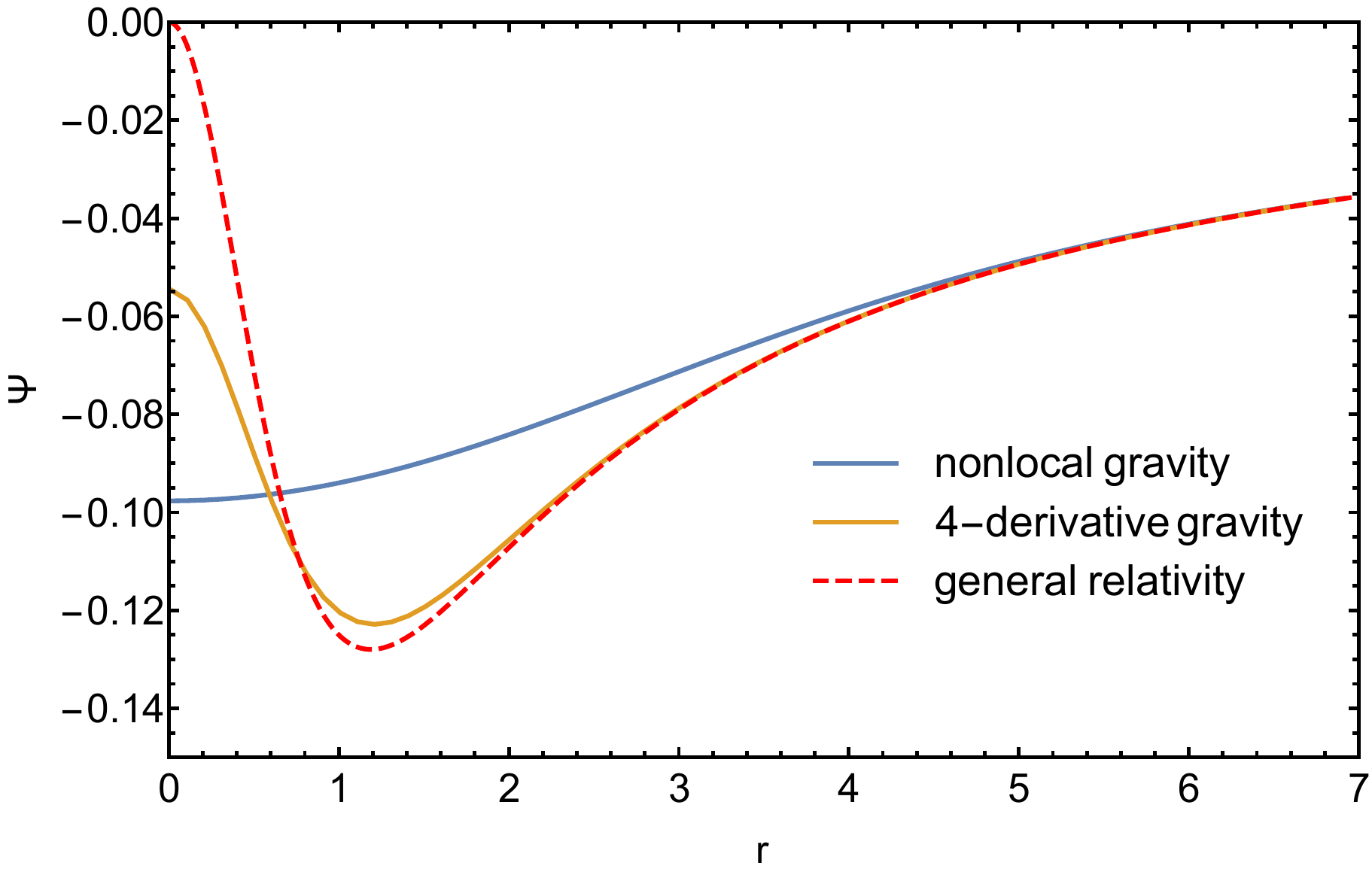}
	}
	\protect\caption{Behaviors of the metric potentials (a) $\Phi$ and (b) $\Psi$ sourced by a scalar field with polynomial potential in Eq.~\eqref{polyn-pot}. The blue solid line corresponds to the nonlocal gravity case with $f(\nabla^2)=e^{-\nabla^2/\mu^2};$ the orange solid line to the four-derivative case with $f(\nabla^2)=1-\alpha\nabla^2/2;$ whereas the red dashed line to the GR case $f(\nabla^2)=1.$ We set $2GM=1/2,$ $R=1,$ $\alpha=0.1$ and $\mu=1.$ 
	}\label{fig2}
\end{figure}

In the case of four-derivative gravity the field equations for the metric potentials are
\begin{eqnarray}
&&\left(1-\frac{\alpha}{2}\nabla^2\right)\nabla^2\Phi(r)=-G\,\frac{64\pi}{3g^2}\frac{r^2-2R^2}{(r^2+R^2)^4}\,,\nonumber\\[2mm]
&&\left(1-\frac{\alpha}{2}\nabla^2\right) \nabla^2\Psi(r)=G\,\frac{64\pi}{3g^2}\frac{2r^2-R^2}{(r^2+R^2)^4}\,,
\label{field-eq-pot-polyn-four}
\end{eqnarray}
which can be solved by means of the Fourier transform method. Indeed, by using Eqs.~\eqref{fourier transfor-polyn}, we can recast the metric potentials in terms of the anti-transform integrals:
\begin{eqnarray}
\Phi(r)&=& -\frac{2GMR^2}{\pi r}\int_0^{\infty} {\rm d}k\,\frac{k\,\sin(kr)}{1+\alpha k^2/2}e^{-Rk}\left(1+\frac{1}{Rk}+\frac{1}{R^2k^2}\right)\,,\\[2mm]
\Psi(r)&=&  \frac{2GMR^2}{\pi r}\int_0^{\infty} {\rm d}k\,\frac{k\,\sin(kr)}{1+\alpha k^2/2}e^{-Rk}\left(1-\frac{1}{Rk}-\frac{1}{R^2k^2}\right) \,.
\label{pot-nonlocal-grav-polyn-four}
\end{eqnarray}
The numerical solutions of the integrals in Eqs.~\eqref{pot-nonlocal-grav-polyn-four} are shown in Fig.~\ref{fig2}.

\subsection{Ghost-free nonlocal gravity}

In the nonlocal gravity case the field equations for the metric potentials are
\begin{eqnarray}
&&e^{-\nabla^2/\mu^2}\nabla^2\Phi(r)=-G\,\frac{64\pi}{3g^2}\frac{r^2-2R^2}{(r^2+R^2)^4}\,,\nonumber\\[2mm]
&&e^{-\nabla^2/\mu^2} \nabla^2\Psi(r)=G\,\frac{64\pi}{3g^2}\frac{2r^2-R^2}{(r^2+R^2)^4}\,.
\label{field-eq-pot-IDG-four}
\end{eqnarray}
By using again the Fourier transform method we can write the metric potentials in the following integral form:
\begin{eqnarray}
\Phi(r)&=& -\frac{2GMR^2}{\pi r}\int_0^{\infty} {\rm d}k\,k\,\sin(kr)e^{-Rk-k^2/\mu^2}\left(1+\frac{1}{Rk}+\frac{1}{R^2k^2}\right)\,,\\[2mm]
\Psi(r)&=&  \frac{2GMR^2}{\pi r}\int_0^{\infty} {\rm d}k\,k\,\sin(kr)e^{-Rk-k^2/\mu^2}\left(1-\frac{1}{Rk}-\frac{1}{R^2k^2}\right) \,.
\label{pot-nonlocal-grav-IDG-four}
\end{eqnarray}
We have shown the numerical solutions of the integrals~\eqref{pot-nonlocal-grav-IDG-four} in Fig.~\ref{fig2} in comparison with the GR and four-derivative cases. 

From the two plots we can see that the nonsingular scalar field source gives nonsingular metric potentials in all three gravitational theories under investigation. We can also clearly understand that nonlocality, or in other words infinite-order derivatives, weaken gravity more as compared to four-derivative and GR cases. Moreover, nonlocality drastically modifies the short-distance contribution in the relativistic component $\Psi.$ Indeed, in the infinite-derivative gravity case for some values of the nonlocal scale $\mu$ the potentials are always monotonically decreasing when going from $r=\infty$ to $r=0.$ This might mean that in the nonlocal gravity case possible repulsive contributions at small distances can be avoided. However, this effect is only true for sufficiently small values of $\mu,$  whereas for large $\mu$ the monotonicity is lost consistently with the fact that for $\mu\rightarrow \infty$ GR result must be recovered.

\section{Tachyon potential in open string field theory}\label{sec-tachyon}

In this section we consider the tachyon potential in open string field theory~\cite{Gerasimov:2000zp}, namely the following (flat) Lagrangian: 
\begin{eqnarray}
\mathcal{L}=\frac{1}{2}\phi\Box\phi+m_s^2\phi^2\log\left(\frac{\phi^2}{m_s^2e}\right)\,,\label{open-tach-lagr}
\end{eqnarray}
where $m_s$ is the string mass scale.

The scalar field equation in Eq.~\eqref{lin-field-scalar} now reads
\begin{eqnarray}
\left(\partial_r^2+\frac{2}{r}\partial_r\right)\phi=-2m_s^2\phi\log\left(\frac{\phi^2}{m_s^2}\right) \,,
\end{eqnarray}
and it admits the following exact solution
\begin{eqnarray}
\phi(r)=e^{3/2}m_se^{-m_s^2r^2}\,,
\end{eqnarray}
where $R=1/m_s$ can be interpreted as the radius of the lump. 

The components of the stress-energy tensor necessary for the computation of the linearized spacetime metric are
\begin{eqnarray}
T_{00}(r)&=&2e^3m^4_se^{-2m^2r^2}(2m^2_sr^2-1)\,,\\
T(r)+2T_{00}(r)&=&4e^3m^4_se^{-2m^2_sr^2}(1-m^2_sr^2)\,.
\end{eqnarray}
The mass-energy $M$ of such a lump configuration is given by
\begin{eqnarray}
M=\int {\rm d}^3r\,T_{00}(r)=\frac{e^3\pi^{3/2}m_s}{2\sqrt{2}}\,.\label{mass-open}
\end{eqnarray}
The corresponding Schwarzschild radius of the lump is
\begin{eqnarray}
R_{\rm sch}=2GM=\frac{e^3\pi^{3/2}Gm_s}{\sqrt{2}}\,.
\end{eqnarray}
Therefore, our weak-field approximation remains valid as long $R=1/m_s\gtrsim R_{\rm sch},$ which imposes the following constraint:
\begin{eqnarray}
m_s^2\lesssim \left(  \frac{e^3\pi^{3/2}}{\sqrt{2}}  \right)M_p^2\,,\label{ineq-open}
\end{eqnarray}
where we have introduced the Planck mass $M_p=1/\sqrt{G}.$ Hence, for a string mass $m_s$ comparable in magnitude with the Planck mass $M_p$ we can hit a nonlinear regime dominated by black-hole formation.

Let us now compute the linearized spacetime metric potentials for the three different theories of gravity as done for the free massive scalar field and the polynomial potentials in the previous sections.

\subsection{General relativity}

In the GR case ($f=1$) the field equations for the metric potentials are
\begin{eqnarray}
&&\nabla^2\Phi(r)= 16\pi G\,e^3m^4_se^{-2m_s^2r^2}(1-m_s^2r^2)\,,\nonumber\\[2mm]
&& \nabla^2\Psi(r)=8\pi G\,e^3m^4_se^{-2m_s^2r^2}(2m_s^2r^2-1)\,.
\label{field-eq-pot-open}
\end{eqnarray}
By imposing asymptotic flatness and setting the homogeneous solution to zero as an extra boundary condition, and using the following Fourier transforms 
\begin{eqnarray}
\int {\rm d}^3r\,  e^{-2m_s^2r^2}(1-m_s^2r^2)e^{-i\vec{k}\cdot\vec{r}}&=& -\frac{\pi^{3/2}e^{-k^2/8m_s^2}(k^2-4m_s^2)}{16\sqrt{2}m_s^5}  \,,\nonumber\\[2mm]
\int {\rm d}^3r\, e^{-2m^2_sr^2}(2m^2_sr^2-1)e^{-i\vec{k}\cdot\vec{r}}&=& \frac{\pi^{3/2}e^{-k^2/8m_s^2}(k^2+4m_s^2)}{32\sqrt{2}m_s^5}   \,,
\label{fourier transfor-open}
\end{eqnarray}
we can obtain the solutions
\begin{eqnarray}
&& \Phi(r)=  -\frac{GM}{r}\left({\rm Erf}\left(\sqrt{2}m_s r\right)+2\sqrt{\frac{2}{\pi}}m_s r e^{-2m_s^2r^2}\right)\,,\\[2mm]
&& \Psi(r)=-\frac{GM}{r}\left({\rm Erf}\left(\sqrt{2}m_s r\right)-2\sqrt{\frac{2}{\pi}}m_s r e^{-2m_s^2r^2}\right)\,,
\end{eqnarray}
where $M$ is mass-energy of the lump computed in Eq.~\eqref{mass-open}. In the large-distance limit, we consistently recover the Newtonian $1/r$ fall-off
\begin{eqnarray}
r\gg R \quad \Rightarrow \quad \Phi,\Psi \sim -\frac{GM}{r}\,.
\end{eqnarray}
Whereas, in the short-distance regime we have
\begin{eqnarray}
r\rightarrow 0 \quad \Rightarrow \quad \Phi \rightarrow -4\sqrt{\frac{2}{\pi}}GM m_s,\quad \Psi\rightarrow 0\,.
\end{eqnarray}
From the last expression, we understand that the linear regime can be trusted as long as  $1/m_s>8\sqrt{2/\pi}GM ,$ which is consistent with Eq.~\eqref{ineq-open}, namely with the fact that lump's radius $R=1/m_s$ must be larger than the corresponding Schwarzschild radius.

The behavior of the metric potentials is shown in Fig.~\ref{fig3}.

\subsection{Four-derivative gravity}

\begin{figure}[t!]
	\centering
	\subfloat[Subfigure 1 list of figures text][]{
		\includegraphics[scale=0.428]{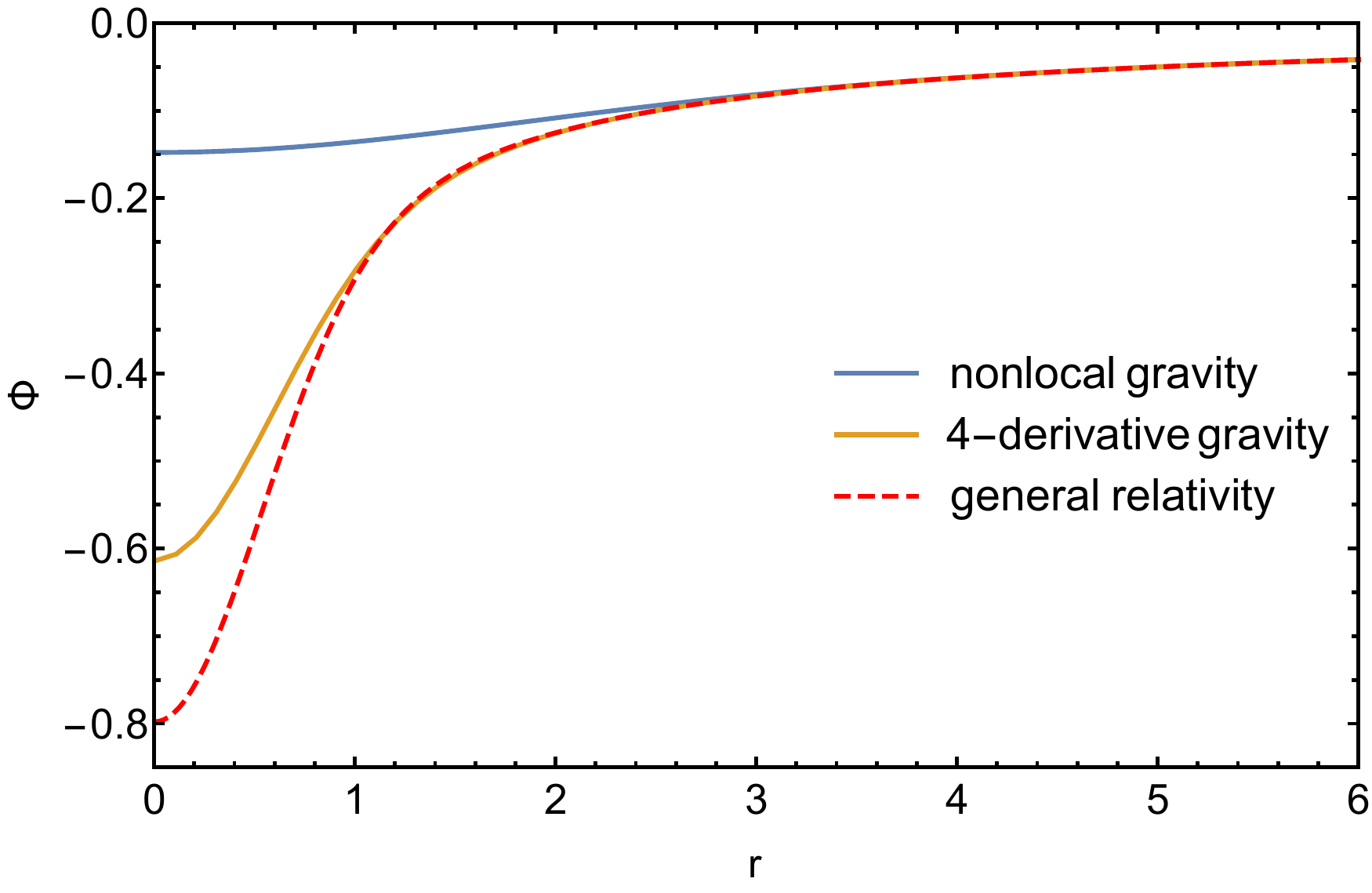}}
	\subfloat[Subfigure 2 list of figures text][]{
		\includegraphics[scale=0.428]{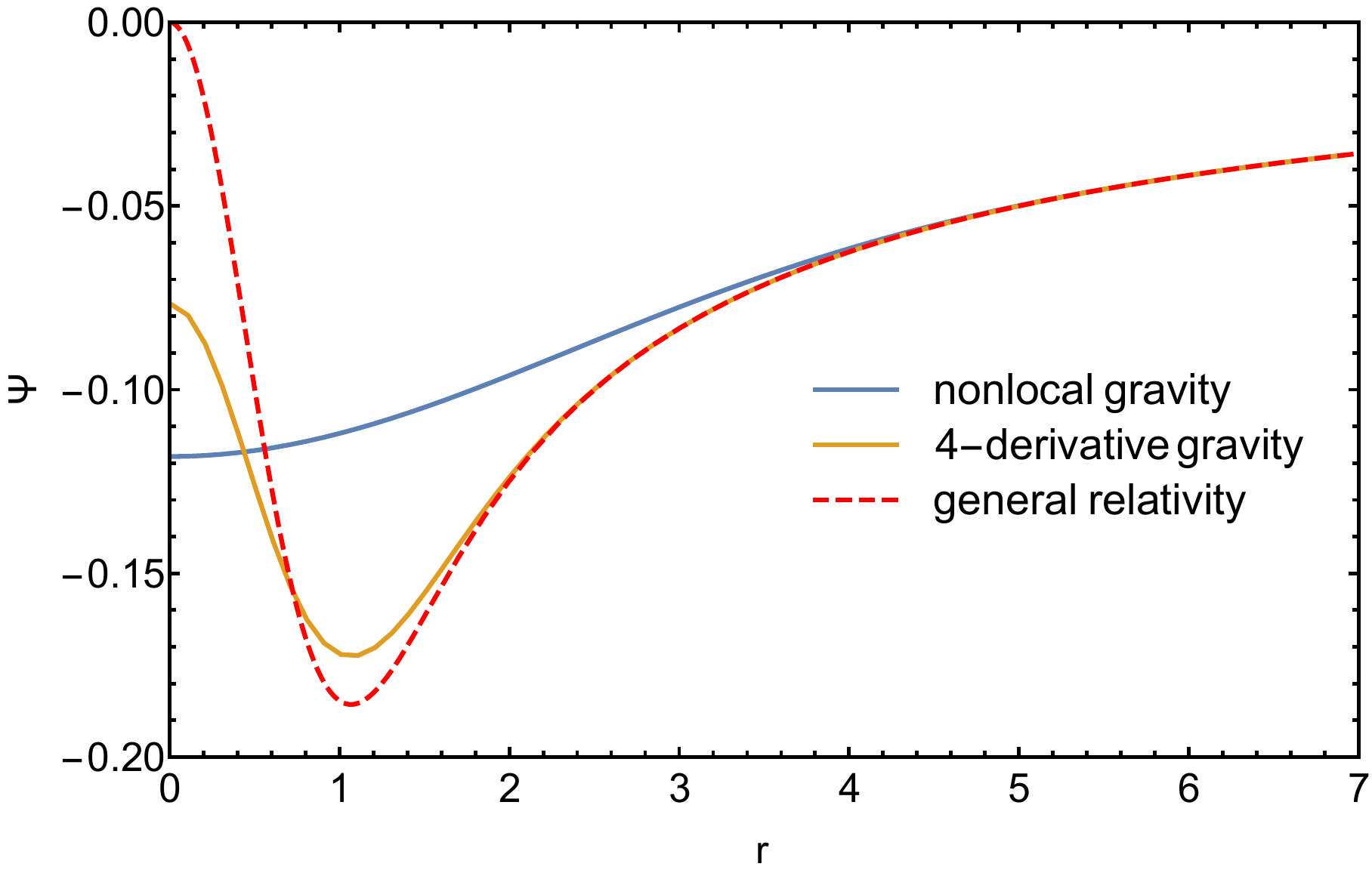}
	}
	\protect\caption{Behaviors of the metric potentials (a) $\Phi$ and (b) $\Psi$ sourced by a tachyon in open string field theory; see Eq.~\eqref{open-tach-lagr}. The blue solid line corresponds to the nonlocal gravity case with $f(\nabla^2)=e^{-\nabla^2/\mu^2};$ the orange solid line to the four-derivative case with $f(\nabla^2)=1-\alpha\nabla^2/2;$ whereas the red dashed line to the GR case $f(\nabla^2)=1.$ We set $2GM=1/4,$ $m_s=1,$ $\alpha=0.1$ and $\mu=1.$ 
	}\label{fig3}
\end{figure}

In the case of four-derivative gravity the field equations for the metric potentials are
\begin{eqnarray}
&&\left(1-\frac{\alpha}{2}\nabla^2\right)\nabla^2\Phi(r)=16\pi G\,e^3m^4_se^{-2m^2_sr^2}(1-m^2_sr^2)\,,\nonumber\\[2mm]
&&\left(1-\frac{\alpha}{2}\nabla^2\right) \nabla^2\Psi(r)=8\pi G\,e^3m^4_se^{-2m^2_sr^2}(2m^2_sr^2-1)\,.
\label{field-eq-pot-open-four}
\end{eqnarray}
By using the following Fourier transforms in Eq.~\eqref{fourier transfor-open}, we can recast the metric potentials in terms of the anti-transform integrals:
\begin{eqnarray}
\Phi(r)&=& -\frac{GM}{2\pi m_s^2 r}\int_0^{\infty} {\rm d}k\,\frac{k\,\sin(kr)}{1+\alpha k^2/2}e^{-k^2/8m_s^2}\left(1+\frac{4m_s^2}{k^2}\right)\,,\\[2mm]
\Psi(r)&=&  \frac{GM}{2\pi m_s^2 r}\int_0^{\infty} {\rm d}k\,\frac{k\,\sin(kr)}{1+\alpha k^2/2}e^{-k^2/8m_s^2}\left(1-\frac{4m_s^2}{k^2}\right) \,.
\label{pot-nonlocal-grav-open-four}
\end{eqnarray}
The numerical solutions of the integrals in Eqs.~\eqref{pot-nonlocal-grav-open-four} are shown in Fig.~\ref{fig3}.

\subsection{Ghost-free nonlocal gravity}

In the nonlocal gravity case the field equations for the metric potentials are
\begin{eqnarray}
&&e^{-\nabla^2/\mu^2}\nabla^2\Phi(r)=16\pi G\,e^3m^4_se^{-2m^2_sr^2}(1-m^2_sr^2)\,,\nonumber\\[2mm]
&&e^{-\nabla^2/\mu^2} \nabla^2\Psi(r)=8\pi G\,e^3m^4_se^{-2m^2_sr^2}(2m^2_sr^2-1)\,.
\label{field-eq-pot-IDG-open}
\end{eqnarray}
By using again the Fourier transform method we can write the metric potentials in the following integral form:
\begin{eqnarray}
\Phi(r)&=& -\frac{GM}{2\pi m_s^2 r}\int_0^{\infty} {\rm d}k\,k\,\sin(kr)e^{-k^2(1/\mu^2+1/8m_s^2)}\left(1+\frac{4m_s^2}{k^2}\right)\,,\\[2mm]
\Psi(r)&=&  \frac{GM}{2\pi m_s^2 r}\int_0^{\infty} {\rm d}k\,k\,\sin(kr)e^{-k^2(1/\mu^2+1/8m_s^2)}\left(1-\frac{4m_s^2}{k^2}\right)\,.
\label{pot-nonlocal-grav-IDG-open}
\end{eqnarray}
Fortunately, in this case the integrals can be solved analytically and give
\begin{eqnarray}
\Phi(r)&=& -\frac{GM}{r}\left({\rm Erf}\left(\frac{\sqrt{2}m_s\mu r}{\sqrt{\mu^2+8m_s^2}}\right)+2\sqrt{\frac{2}{\pi}}\frac{m_s\mu^3 r}{(\mu^2+8m_s^2)^{3/2}} e^{-\frac{2m_s^2\mu^2}{\mu^2+8m_s^2}r^2}\right)\,,\label{potgrav-IDG-open-analytic-phi}\\[2mm]
\Psi(r)&=&  -\frac{GM}{r}\left({\rm Erf}\left(\frac{\sqrt{2}m_s\mu r}{\sqrt{\mu^2+8m_s^2}}\right)-2\sqrt{\frac{2}{\pi}}\frac{m_s\mu^3 r}{(\mu^2+8m_s^2)^{3/2}} e^{-\frac{2m_s^2\mu^2}{\mu^2+8m_s^2}r^2}\right)\,.
\label{potgrav-IDG-open-analytic-psi}
\end{eqnarray}
In the short-distance regime, the two metric potentials assume the following values
\begin{eqnarray}
r\rightarrow 0 \quad \Rightarrow \quad \Phi \rightarrow -4\sqrt{\frac{2}{\pi}}GM m_s\frac{\mu(\mu^2+4m_s^2)}{(\mu^2+8m_s^2)^{3/2}}\,,\quad \Psi\rightarrow -16\sqrt{\frac{2}{\pi}}\frac{GMm_s^3\mu}{(\mu^2+8m_s^2)^{3/2}}\,.
\end{eqnarray}
We have shown the behavior of the metric potentials~\eqref{potgrav-IDG-open-analytic-phi} and~\eqref{potgrav-IDG-open-analytic-psi} in Fig.~\ref{fig3} in comparison with the GR and four-derivative cases. 

From Fig.~\ref{fig3}  it is clear that the regularized nature of the source implies that also the metric potentials are regular in all the gravitational theories under investigation. We notice again that the presence of infinite-order derivatives can drastically change the short-distance behavior of the gravitational interaction by avoiding possible repulsive forces, indeed both $\Phi$ and $\Psi$ are monotonic functions. Similarly to the case of polynomial scalar field potential, such a nonlocal effect is lost for sufficiently large values of $\mu.$

Since we have obtained analytic solutions for the gravitational potentials, we can explicitly find the range of values of $\mu$ for which $\Psi$ is monotonic. We can compute the derivative $\Psi^\prime={\rm d}\Psi/{\rm d}r$ and check its sign in the short-distance regime. If $\Psi^\prime>0,$ then the potential is monotonic; whereas if $\Psi^\prime<0,$ then monotonicity is lost.

By computing the derivative, we get
\begin{eqnarray}
\Psi^\prime (r\simeq 0)= 16GMm_s^3\sqrt{\frac{2}{\pi}}\frac{\mu^3(4m_s^2-\mu^2)r}{3(\mu^2+8m_s^2)^{5/2}}+\cdots\,,
\end{eqnarray}
where the dots stand for higher-order contributions. We understand that the potential $\Psi$ is monotonic when $\mu\leq 2m_s;$ whereas in the opposite regime $\mu>2m_s$ monotonicity is lost and possible repulsive contributions can manifest.

\section{Discussion and conclusions}\label{conclus}

In this paper we studied spacetime metrics sourced by scalar lumps in Einstein's GR, in four-derivative gravity and in ghost-free nonlocal gravity. We considered three type of scalar fields: a free massive scalar field as a warm up exercise, and then a polynomial potential and the tachyon in open string field theory as more physical configurations. We noticed that unlike in GR and four-derivative gravity, in the case of infinite-derivative gravity the nonlocality not only weakens the gravitational interaction but it is also able to avoid possible repulsive contributions for sufficiently small value of the nonlocal energy scale. It is worth to mention that a similar feature can be also obtained in theories of gravity with derivative-order higher than four, e.g. in sixth order gravity.

Let us emphasize that we worked in the linearized regime in which the metric potentials had to satisfy the inequalities $2|\Phi|,2|\Psi|<1.$ Although we were in a weak-field scenario, the metric solutions we found for the polynomial and tachyon scalar potentials are valid all the way from $r=\infty$ up to $r=0$ as long as the aforementioned inequalities hold true. In this sense, our solutions can describe \textit{horizonless} compact objects made up of scalar fields. 

We can define the compactness of an astrophysical object as $\mathcal{C}\equiv GM/\mathcal{R},$ where $\mathcal{R}$ is an effective radius within which the mass-energy $M$ is localized. For black holes (of radius $\mathcal{R}=2GM$) we have $\mathcal{C}_{\rm BH}=1/2,$ whereas for any horizonless object $\mathcal{C}<1/2.$ As for the nonsingular spacetime metric solutions found in Sections~\ref{sec-polyn scalar} and~\ref{sec-tachyon}, in GR the effective radius $\mathcal{R}$ coincides with the physical radius of the source, but in higher-derivative theories in general it does not. Indeed, in the four-derivative and nonlocal cases the compactness also depends on the parameters $\alpha$ and $\mu,$ respectively, which enter in the definition of $\mathcal{R}.$

In fact, more generally the compactness is proportional to the value of the potential $\Phi$ at the origin, i.e. $\mathcal{C}\propto \Phi(0).$ From this last observation it follows that higher-order derivatives can make the corresponding gravitational system less compact; see also Figs.~\ref{fig2} and~\ref{fig3}. In particular, nonlocality (i.e. infinite-order derivatives) give the less compact configuration as compared to the four-derivative and GR cases.

In the recent years, since the first observation of gravitational waves~\cite{Abbott:2016blz}, there have been lots of new investigations on compact objects in theories beyond Einstein's GR. One of the key feature of many exotic astrophysical objects is that they are horizonless. In particular, ultra-compact objects which still possess a photon sphere can generate \textit{echoes} during catastrophic events like binary mergers. See Ref.~\cite{Cardoso:2019rvt} for a very recent review on theoretical and phenomenological aspects of exotic compact objects.  

Hence, as a future investigation it will be very interesting to study some phenomenology of horizonless compact objects sourced by scalar fields in higher (infinite) derivative gravity. In particular, by introducing higher multipole terms and a non-zero angular velocity, and making a stability analysis, one can assume that such objects are the remnant of a binary merger and analyse their relaxation process by computing quasi-normal modes and waveforms~\cite{Maggio:2020jml}. Such a study will allow to test and constrain both gravity and matter sectors.


\subsection*{Acknowledgements}
The authors are grateful to Breno L. Giacchini for enlightening discussions and useful comments. L.~B. acknowledges financial support from JSPS and KAKENHI Grant--in--Aid for JSPS Postdoctoral Fellows No.~JP19F19324.



\end{document}